\documentclass[preprint2]{aastex6}
\shorttitle{Local Heating Events in Flux Emergence}
\shortauthors{Toriumi et al.}
\begin{document}


% \title{Hinode-IRIS Analysis on Local Heating Events in Emerging Flux Regions}
% \title{Various Local Heating Events in the Earliest Phase of Emerging Flux Regions}
\title{Various Local Heating Events in the Earliest Phase of Flux Emergence}

%% Use \author, \affil, plus the \and command to format author and affiliation 
%% information.  If done correctly the peer review system will be able to
%% automatically put the author and affiliation information from the manuscript
%% and save the corresponding author the trouble of entering it by hand.
%%
%% The \affil should be used to document primary affiliations and the
%% \altaffil should be used for secondary affiliations, titles, or email.

%% Authors with the same affiliation can be grouped in a single

\author{Shin Toriumi$^{1,}$\altaffilmark{2}, Yukio Katsukawa$^{1}$, Mark C. M. Cheung$^{1, 3}$}
\affil{$^{1}$National Astronomical Observatory of Japan, 2-21-1 Osawa, Mitaka, Tokyo 181-8588, Japan\\
$^{3}$Lockheed Martin Solar and Astrophysics Laboratory, 3251 Hanover Street, Building/252, Palo Alto, CA 94304, USA}

%% Use the \and command so offset the last author.
% \and

% \author{Jeff Lewandowski\altaffilmark{5}}
% \affil{IOP Publishing, Washington, DC 20005}


\altaffiltext{2}{shin.toriumi@nao.ac.jp}

\begin{abstract}
Emerging flux regions (EFRs)
are known to exhibit
various sporadic local heating events
in the lower atmosphere.
To investigate the characteristics
of these events,
especially to link
the photospheric magnetic fields
and atmospheric dynamics,
we analyze {\it Hinode}, {\it IRIS},
and {\it SDO} data
of a new EFR
in NOAA AR 12401.
Out of 151 bright points (BPs)
identified in {\it Hinode}/SOT Ca images,
29 are overlapped
by an SOT/SP scan.
Seven BPs in the EFR center
possess mixed-polarity magnetic backgrounds
in the photosphere.
Their {\it IRIS} UV spectra
(e.g., \ion{Si}{4} 1402.8 {\AA})
are strongly enhanced
and red- or blue-shifted
with tails reaching $\pm 150\ {\rm km\ s}^{-1}$,
which is highly suggestive
of bi-directional jets,
and each brightening lasts
for 10 -- 15 minutes
leaving flare-like light curves.
Most of this group
show bald patches,
the U-shaped photospheric magnetic loops.
Another 10 BPs are found
in unipolar regions
at the EFR edges.
They are generally weaker
in UV intensities
and exhibit systematic redshifts
with Doppler speeds
up to $40\ {\rm km\ s}^{-1}$,
which could exceed
the local sound speed
in the transition region.
Both types of BPs
show signs of
strong temperature increase
in the low chromosphere.
These observational results
support the physical picture
that heating events
in the EFR center
are due to magnetic reconnection
within cancelling undular fields
like Ellerman bombs,
while the peripheral heating events
are due to shocks
or strong compressions
caused by
fast downflows
along the overlying arch filament system.
\end{abstract}




\section{Introduction} \label{sec:introduction}

Magnetic flux emergence
and the formation of active regions (ARs) is
one of the most fascinating phenomena
of the Sun
\citep{zwa85,che14,sch14}.
Apart from catastrophic energy-releasing phenomena
such as flares and CMEs
\citep{shi11}
and X-ray jets
observed higher in the corona
\citep{shi92},
a wide variety of
smaller-scale events
are found in ARs.
Repeated brightenings
and jet ejections occur
in light bridge structures
\citep{roy73,asa01,tor15a,tor15b},
whereas jet-like features are detected
in sunspot penumbrae
\citep{kat07,vis15a}.
Coupling of magnetic fields
and local convection
may drive such activity
\citep{spr06}.

Even at the earliest stage
of flux emergence,
various small-scale events
are observed.
A commonly reported type
of small-scale events is
Ellerman bombs,
brightenings in the H$\alpha$ line wings
(\citealt{ell17}; see review by \citealt{rut13}).
They have a typical size of $\lesssim 1\arcsec$
with a lifetime of $\lesssim 10$ minutes
\citep[e.g.,][]{geo02}.
As shown by radiative transfer models
\citep[e.g.,][]{kit83},
the confinement of the intensity enhancement
to the wings
of the H$\alpha$ suggests that
Ellerman bombs are heating events
in the photosphere,
while high-resolution observations revealed that
the Ellerman bombs are located at the sites
of magnetic reconnection
in the photosphere
with upright flame-like structures
\citep{wat11,vis13,vis15b}.
Recent observations by
{\it Interface Region Imaging Spectrograph}
({\it IRIS}; \citealt{dep14})
show that
the location of the UV bursts
({\it IRIS} bombs)
are coincident with locations
where opposite-polarity magnetic patches
in an emerging flux meet
\citep{pet14,tia16}.
Independently,
numerical models of AR formation
following flux emergence
point to the importance
of reconnection
in opposite-polarity patches
\citep{che10}.
Such local heating events occur
anywhere in the EFRs
with dynamic temporal variations.

Therefore,
in order to reveal the nature of these events,
it is crucially important
to focus on its earliest phase
of the flux emergence
(within a few hours
from the initial appearance),
detect as many events as possible
with a wide field-of-view (FOV)
as well as
sufficient temporal and spatial resolutions,
and analyze them in a statistical way.
Furthermore,
to investigate the physical connection
between photospheric magnetic fields
and atmospheric dynamics,
we need simultaneous photospheric spectropolarimetry
and chromospheric and transition-region spectrograph
on the EFRs,
which can only be achieved
through the coordinated observation
by {\it Hinode} \citep{kos07}
and {\it IRIS}.

Although this kind of observation
is difficult and thus is still rare,
here we report on
the analysis of
a data set
that satisfies
all the afore-mentioned conditions.
The rest of this paper
is organized as follows.
First, in Section \ref{sec:observations},
we describe the observations.
Then,
in Section \ref{sec:results},
we present the analysis results,
which are summarized
in Section \ref{sec:summary}.
Finally,
in Section \ref{sec:discussion},
we discuss the observed phenomena
and possible physical pictures.

\section{Observations}\label{sec:observations}

The target EFR of this study appeared
within the pre-existing AR NOAA 12401.
This AR had a simple bipole structure,
and the target EFR started emergence
around 06:00 UT,
2015 August 19,
at the center of the AR.
The magnetic flux increased rapidly
from around 09:00 UT,
which was captured
by {\it Hinode} and {\it IRIS}
(coordinated observation).
At this time,
the target EFR was located
around $(320\arcsec, -270\arcsec)$
on the solar disk,
i.e., at an angle
of $\sim 26^{\circ}$
from the disk center (DC).

The spectropolarimeter
(SP; \citealt{lit13})
of the {\it Hinode}/Solar Optical Telescope (SOT; \citealt{tsu08})
obtained a single raster scan
from 10:35 to 11:00 UT
in the \ion{Fe}{1} lines
at 6301.5 and 6302.5 {\AA}.
The SP scan has a pixel size along the slit
and a step size
of $0\farcs 32$ and $0\farcs 30$,
respectively,
and the FOV
is $113\farcs 5\times 122\farcs 9$.
Also,
the broadband filter imager
sequentially shot
the \ion{Ca}{2} H (3968.5 {\AA}) images
with a temporal gap
between 08:29 and 10:34 UT.
The \ion{Ca}{2} data have
a pixel size of $0\farcs 22\times 0\farcs 22$,
FOV of $223\farcs 2\times 111\farcs 6$,
and cadence of 63 s.
We applied standard procedures
for prepping the SOT data sets.
The photospheric vector magnetic fields
were obtained from the level 2 SP data
by using MERLIN \citep{lit07}
and AZAM \citep{lit95}.

The {\it IRIS} data
were made between 07:48 and 10:54 UT.
This observation sequence
(OBS 3860106092)
consists of large sparse 64-step rasters
and slit-jaw images (SJIs).
During this period,
34 raster scans were made
for \ion{C}{2} 1334.5 and 1335.7 {\AA},
\ion{Si}{4} 1402.8 {\AA},
and \ion{Mg}{2} h 2803.5 and k 2796.4 {\AA},
which have
a step size, step cadence, and raster cadence
of $1\farcs 0$, 5.2 s, and 330 s,
respectively.
The pixel size is $0\farcs 33$ along the slit
and the FOV of each scan
is $63\farcs 9\times 118\farcs 8$.
The SJIs are composed of
1330, 1400, 2796, and 2832 {\AA} filtergrams,
each having a pixel size of $0\farcs 33\times 0\farcs 33$,
FOV of $119\arcsec\times 118\arcsec$,
and cadence of 20.6 s.
We used the level 2 data,
in which the dark-current subtraction,
flat fielding,
and geometrical
and orbital variation corrections
were taken into account\footnote{Absolute wavelength calibration
was done using the cold (photospheric)
\ion{Ni}{1} 2799.5 {\AA} line.}.

We also analyzed the data taken
by the {\it Solar Dynamics Observatory}
({\it SDO}; \citealt{pes12}).
The photospheric evolution 
was monitored
by the Helioseismic and Magnetic Imager
(HMI; \citealt{sche12,scho12}),
while the response in the higher atmosphere
was captured
by the Atmospheric Imaging Assembly
(AIA; \citealt{lem12}).
The pixel sizes
for the HMI and AIA images
are $0\farcs 5$ and $0\farcs 6$,
respectively.
The time cadence is
45 s for the HMI images
and 24 s for the AIA's two ultraviolet images
(1600 and 1700 {\AA}).
The center-to-limb variation
was first subtracted
from the AIA images
by applying the method
in \citet{tor14}.
All the observational data
were co-aligned
by using the AIA images
as a reference.

\section{Results}\label{sec:results}

\subsection{Overall Evolution}

Figure \ref{fig:overview} shows
the overview of the target EFR
in this study:
see movie for the temporal evolution.
In the center of the $30\arcsec\times 30\arcsec$ FOV,
the HMI magnetogram shows that
a number of tiny magnetic elements
of positive and negative polarities
(a few arcsec)
emerge at the surface
and separated each other
in the northeast-southwest orientation.
Eventually,
the magnetic elements
merge together to form pores
at the edges of the EFR
(see the HMI intensitygram:
pore size $\sim 5\arcsec$).

At the same time,
in the chromosphere,
various brightening events
are seen in different filtergrams
(1400, 1600, and 1700 {\AA}).
They are located
mostly in the central region
with mixed magnetic polarities
but also in the peripheral
unipolar regions.
The bright structures are seen to be larger
in 1600 and 1700 {\AA} images
not only because of the large pixel sizes
but also of the scattering effects
\citep[see, e.g.,][]{rut16}.
Other filtergrams
(e.g., 1330 and 2796 {\AA})
exhibit similar brightening structures
(not shown).

In the map of the \ion{Mg}{2} k
absorption line core (k3),
an arch filament system (AFS)
is clearly seen,
which has an area
of $\sim 30\arcsec\times 20\arcsec$
with a typical individual length
of $\sim 20\arcsec$
and a width
of $\sim 2\arcsec$.
The Doppler velocity map
obtained from the \ion{Si}{4} spectrum
reveals a general trend
that the EFR center
(crests of the AFS)
shows upflows
and the peripheries
(footpoints of the AFS)
have downflows.
These observational results are consistent
with
the H$\alpha$ observations
reported by
\citet{bru67} and \citet{bru69}.
The intensity of the \ion{Mg}{2} triplet line
between \ion{Mg}{2} h and k
is enhanced
at the footpoints
of the AFS.

\subsection{Selection of Brightening Events}\label{sec:selection}

For surveying the variety of
local heating events,
we identified bright points (BPs)
from the SOT \ion{Ca}{2} H images.
The intensity threshold
for picking up such events
is the 5$\sigma$ or higher above the mean
of the AIA 1700 {\AA} quiet-Sun values,
which is suggested by \citet{rut13}.

Figure \ref{fig:eb_sp}(a) shows
a scatter diagram of 
1700 {\AA} intensity
and \ion{Ca}{2} H intensity.
By fitting a straight line
to this doubly logarithmic plot,
we obtained the relation
between the two data sets
and found that the 5$\sigma +$mean criterion
of the 1700 {\AA} images,
which is 4232.1 DN
in the present data set,
is equivalent to
1005.1 DN
in the Ca images.
We adopted this value
for detecting
the BPs in the Ca images
and this criterion led to a total of
151 BPs
from the sequential Ca data series
in the period
from 10:34 to 10:59 UT
(25 frames).

Out of this 151 BPs,
we found that 29 BPs were covered
by the SOT SP scan.
Figure \ref{fig:eb_sp}(b) shows
the distribution
of the 29 BPs
on the SOT SP circular polarization map
(Stokes-V/I signal
at 6301.43 {\AA}),
which represents
the line-of-sight (LOS) component
of the photospheric magnetic fields.
Here,
the BPs have a size
of sub-arcsec to a few arcsec
and are mostly
seen in the magnetized regions.

We then checked the magnetic context
of the 29 BPs
using the degree of mixed polarity
\citep{kat05},
\begin{eqnarray}
  p_{\rm mix}=1-
  \left|
    \frac{
      |\Phi_{+}|-|\Phi_{-}|
    }{
      |\Phi_{+}|+|\Phi_{-}|
    }
  \right|,
\end{eqnarray}
where $\Phi_{\pm}$ is the total LOS flux
of the positive ($+$) or negative ($-$) polarities
within each BP patch.
If $p_{\rm mix}=0$
the BP has
either a positive or negative polarity
(unipolar),
while $p_{\rm mix}\ne 0$ indicates
that the BP has both polarities
within the Ca intensity contour
(mixed-polarity).
As a result,
we found that there are
seven mixed-polarity BPs
and 22 unipolar BPs.

\subsection{Spectral and Polarimetric Features}\label{sec:spectral}

The 29 Ca BPs
identified in Section \ref{sec:selection}
are classified
into the following categories.
The first group is the mixed-polarity events
(seven BPs).
They are distributed
mostly in the center
of the target EFR.
The second and the third groups are
the unipolar events
in the limb-side,
i.e., the south-western end
of the EFR
(five BPs)
and in the DC-side,
i.e., the north-eastern end
(five BPs),
respectively.

There are some remaining unipolar events,
which are distributed in
the EFR center (three BPs)
and the plage regions
outside the EFR
(nine BPs).
The general trends
of these events
are similar to
those of the limb-side
and DC-side unipolar events.

In this section,
we introduce the spectral and polarimetric features
of the mixed-polarity,
limb-side unipolar,
and DC-side unipolar events
with showing the representative BPs.

\subsubsection{Mixed-polarity Events}\label{sec:mix}

The mixed-polarity BPs
are basically located
in the central region
of the EFR.
A typical example
of the mixed-polarity event
is shown in Figure \ref{fig:mix}.
This figure consists of
three
filtergrams
and a circular polarization map
of a BP located at
$(X, Y)=(336\arcsec, -274\arcsec)$
in Figure \ref{fig:eb_sp}(b)
plus the {\it IRIS}
spectra
and SOT SP Stokes-V/I profiles
for the three different locations
within the BP.

Typically, in magnetized regions,
Stokes-V/I
has an antisymmetric profile
with respect to the line center
with two lobes
of opposite signs
in the red and blue wings.
The three Stokes-V/I profiles
in Figure \ref{fig:mix}
show a reversal of the sign,
which indicates the transition
of the LOS magnetic fields
between positive (colored blue)
and negative (red) polarities.
At the central position
of the BP
located between the opposite polarities,
the purple Stokes-V/I profile
shows an asymmetric, irregular shape.
Such shapes are often interpreted
as a co-existence
of closely neighboring two different magnetic components,
e.g., positive and negative polarities,
within a single pixel
\citep{che08},
or as a complex stratification
of magnetized atmosphere
along the LOS direction.

One of the clearest characteristics
of the mixed-polarity BPs
is the strong enhancements
of the
line intensities
and spectral widths.
In Figure \ref{fig:mix},
the spectral lines
have their maximum intensities
at location indicated by
the purple cross,
i.e., at the BP center
(see filtergrams and spectra).
Among others,
the peak intensity
of \ion{Si}{4} spectrum
is as much as 50 times
higher than its quiet-Sun value.
The purple profile at the BP center
shows extended line wings
of Doppler speeds of
up to $\pm 100\ {\rm km\ s}^{-1}$.
Morphologically,
this BP has a size
of about $3\arcsec\times 3\arcsec$
with a point-like bright core.

It is noteworthy here that
the Doppler shifts
of the {\it IRIS} spectra,
especially of the transition-region lines
(\ion{Si}{4} and \ion{C}{2}),
show a positional dependence.
For example,
the spectra indicated
by red color,
i.e., those of the DC side
of the BP,
have red-shifted profiles,
indicating that
the material at the transition-region temperature
is flowing down.
The Doppler velocity
measured at the \ion{Si}{4} peak
is about $20\ {\rm km\ s}^{-1}$.
On the other hand,
the blue-colored spectra,
i.e., the limb-ward ones,
are clearly blue-shifted,
suggesting the upflow
with an absolute velocity
of $\sim 30\ {\rm km\ s}^{-1}$.
Similar Doppler-shift dependency
of the location
(DC-ward or limb-ward)
was recently reported
by \citet{vis15b}.

The \ion{Mg}{2} h and k spectra
often exhibit double-peaked profiles
with the so-called ``self-absorption'' cores.
These central reversals
are caused by the AFS
(see \ion{Mg}{2} k3 image
of Figure \ref{fig:overview}),
indicating that
the Ca BPs are mostly covered
by the thick and dense AFS material
that exerts large opacity.
Similarly,
the \ion{C}{2} spectra
often show central reversals, too,
while the \ion{Si}{4} do not
in this data set.

The subordinate \ion{Mg}{2} triplet line
is sometimes seen in emission.
In Figure \ref{fig:mix},
the triplet emission
is exceptionally remarkable
at the center of the BP
(purple profile).
According to \citet{per15},
the triplet emission
is rare and indicates
a steep temperature increase
above 1500 K
in the lower chromosphere.
These features are also reported
by \citet{vis15b} and \citet{tia16}.

The circular polarization map
in Figure \ref{fig:mix}
also shows the location of
the polarity inversion lines (PILs),
where the magnetic field has
no radial component
(i.e., $B_{z}=0$)
and whether
the magnetic structure
at the PIL is
``dip'' or ``bald patch'',
in which the field lines have
a 
concave-up
configuration,
\begin{eqnarray}
  B_{z}=0\ {\rm and}\ \mbox{\boldmath $B$}\cdot\nabla B_{z}>0,
\end{eqnarray}
or ``top'',
in which the field lines
are
concave down,
\begin{eqnarray}
  B_{z}=0\ {\rm and}\ \mbox{\boldmath $B$}\cdot\nabla B_{z}<0
\end{eqnarray}
\citep{par04,wat08}.
Since we have
only a single-layer magnetogram,
it would be better to explicitly describe
the bald patch as
\begin{eqnarray}
  B_{z}=0\ {\rm and}\ B_{x}
  \frac{\partial B_{z}}{\partial x}
  +B_{y}\frac{\partial B_{z}}{\partial y}>0
\end{eqnarray}
and the counterpart as
\begin{eqnarray}
  B_{z}=0\ {\rm and}\ B_{x}
  \frac{\partial B_{z}}{\partial x}
  +B_{y}\frac{\partial B_{z}}{\partial y}<0.
\end{eqnarray}
Note that we here use
the photospheric vector magnetic fields
$(B_{x}, B_{y}, B_{z})$,
which is transformed
to the local Cartesian reference frame
$(x, y, z)$ with $\hat{\mbox{\boldmath $z$}}$ being
the local radial direction.
It is clearly seen
from the circular polarization map
of Figure \ref{fig:mix} that
the PIL passes through the center
of the Ca BP
and that the magnetic fields have
a bald-patch configuration
(yellow dots).
It is found that,
in five mixed-polarity BPs
out of the seven total events,
the PIL is dominated
by the bald patch:
see Appendix \ref{app:bald}
for PILs of the six remaining mixed-polarity BPs.

\subsubsection{Unipolar Events: Limb-side}\label{sec:lue}

Figure \ref{fig:uni_lm} is
a typical example
of the unipolar BP
in the limb-side
of the target EFR.
This BP is centered at
$(X, Y)=(350\arcsec, -272\arcsec)$
in Figure \ref{fig:eb_sp}(b).
This location is
one of the
the south-western footpoints
of the AFS
(see Figure \ref{fig:overview})
and is surrounded
by a large magnetic patch
of the negative polarity.
In the Stokes-V/I profiles,
the unipolar nature of this BP
is clearly reflected:
they do not show a reversal
or an irregular shape.

The line intensities
are generally weaker
than those of the mixed-polarity events.
For example,
the maximum intensities
of the \ion{Si}{4} and \ion{C}{2} lines
in Figure \ref{fig:uni_lm}
are at most five times
of the quiet-Sun value,
as opposed to much more than 10 times
for the mixed-polarity events.

Another characteristic
of the limb-side unipolar events
is that the \ion{Si}{4} and \ion{C}{2} spectra
coherently show redshifts,
i.e., downflows.
In Figure \ref{fig:uni_lm},
the Doppler velocity
at the peak of the \ion{Si}{4} line
reaches $30\ {\rm km\ s}^{-1}$.

It is also interesting that
the \ion{Mg}{2} lines
do not show regular double-peaked profiles
with
central reversals
but single-peaked
(or top-flat) profiles.
According to \citet{car15},
such single-peaked \ion{Mg}{2} k profiles
are found above the plage regions
and imply the existence of
hot and dense chromosphere of about 6500 K
and a transition region
at a high column mass.
In fact,
the single-peaked \ion{Mg}{2} k profiles
in this data set
are seen mostly in the vicinity of
or above the developing pores.
Their Doppler shifts are not remarkable
compared to those
of the transition-region lines
(\ion{Si}{4} and \ion{C}{2}).
It should also be noted that
the \ion{Mg}{2} triplet emission,
which indicates the lower-chromospheric heating,
are seen
at some locations.

\subsubsection{Unipolar Events: DC-side}

The second group
of the unipolar BPs
is located in the DC-side,
i.e., the north-eastern end,
of the target EFR.
Figure \ref{fig:uni_dc} shows
the sample event around
$(X, Y)=(328\arcsec, -266\arcsec)$
in Figure \ref{fig:eb_sp}(b).

\begin{figure}
  \begin{center}
    \includegraphics[width=170mm]{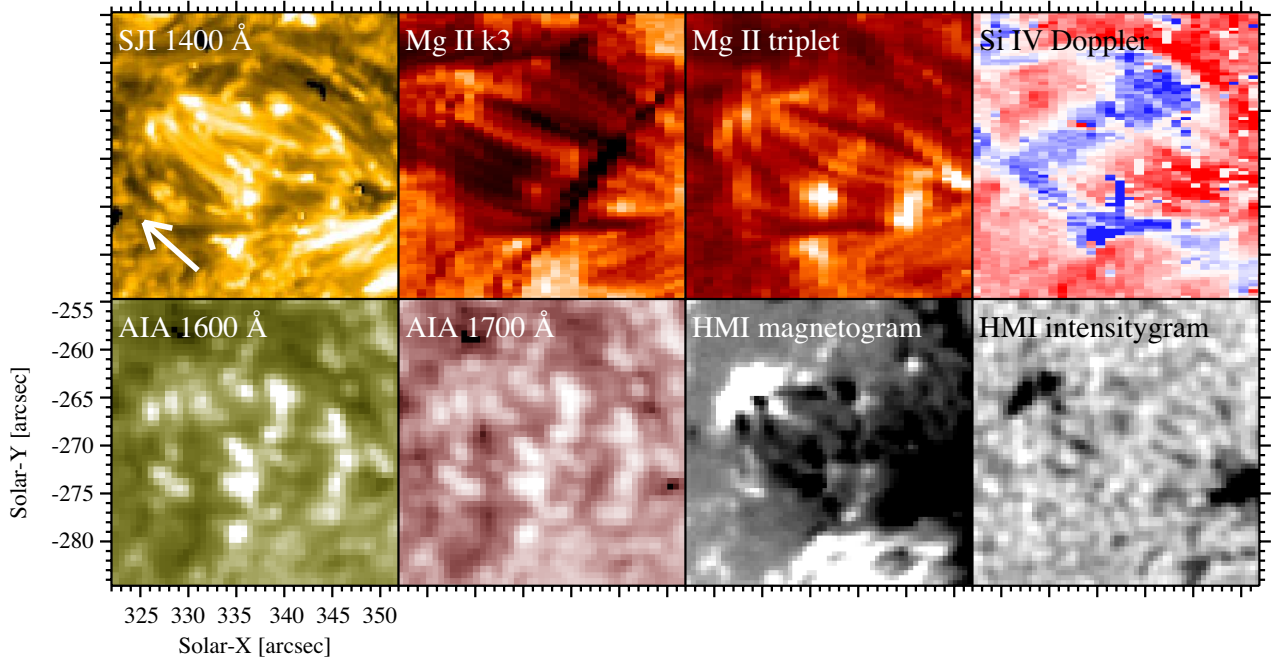}
  \end{center}
  \caption{
    Composite of various observations
    to show the overall evolution
    of the target EFR.
    The corresponding movie
    provides the temporal evolution.
    From top left to bottom right,
    {\it IRIS} SJI 1400 {\AA},
    intensitygram
    at the \ion{Mg}{2} k line core (k3),
    intensitygram
    at the \ion{Mg}{2} triplet line,
    Dopplergram produced from
    the \ion{Si}{4} 1403 {\AA} spectrum
    (blue, white, and red correspond to
    $-10$, $0$, and $+40\ {\rm km\ s}^{-1}$,
    respectively),
    {\it SDO}/AIA 1600 {\AA},
    1700 {\AA},
    {\it SDO}/HMI magnetogram
    (black and white correspond to
    $-200$ and $200\ {\rm G}$,
    respectively),
    and intensitygram.
    The white arrow
    in the top left panel
    indicates the direction
    of the DC.
  }
  \label{fig:overview}
\end{figure}

\begin{figure}
  \begin{center}
    \includegraphics[width=70mm]{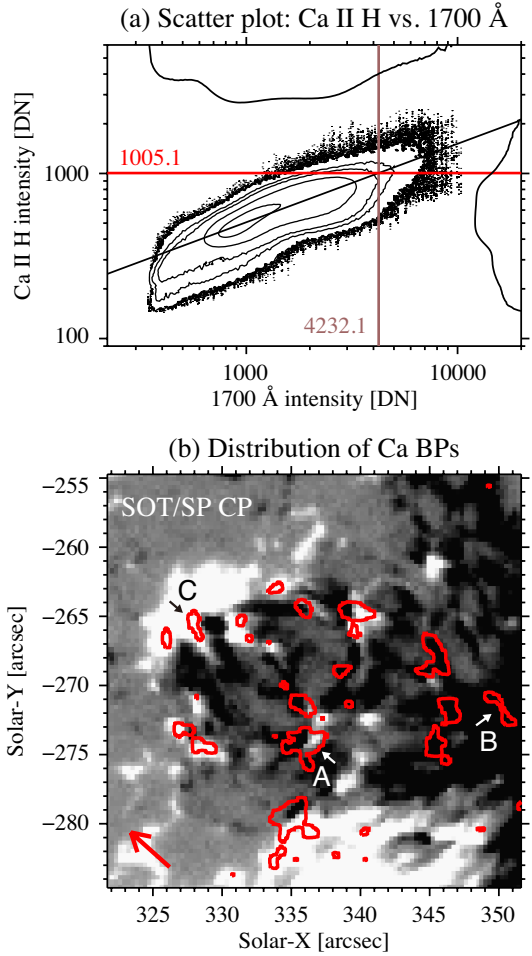}
  \end{center}
  \caption{
    (a) Scatter plot
    of the SOT \ion{Ca}{2} H intensity
    versus the AIA 1700 {\AA} intensity.
    Density contours
    in logarithmic scale
    are plotted
    for large density regions
    to avoid the plot saturation,
    while the logarithmic distributions
    for both parameters
    are shown
    at the right and top
    of the diagram.
    The vertical line indicates
    the $5\sigma+$mean level
    of the 1700 {\AA} intensity
    and the horizontal line
    shows the equivalent value
    for the \ion{Ca}{2} H intensity.
    (b) Distribution of the Ca BPs
    plotted on the SOT SP circular polarization map,
    identifying the 29 BPs
    analyzed in this study.
    Red contour indicates
    the \ion{Ca}{2} H intensity level
    obtained in panel (a),
    while the red arrow at the bottom left
    shows the direction of the DC.
    A, B, and C label
    the three representative BPs
    introduced in Section \ref{sec:spectral}.
  }
  \label{fig:eb_sp}
\end{figure}

This BP also exists
at one footpoint
of the AFS
seen in the \ion{Mg}{2} k3 images
(Figure \ref{fig:overview}).
The Stokes-V/I signals
in Figure \ref{fig:uni_dc}
show that
this BP has a positive magnetic polarity
with no irregular profiles.
The line intensities are again weaker
than those of the mixed-polarity events.
Especially in the 1400 {\AA} map,
the intensity enhancement
is much less marked.

The systematic redshifts are also
seen in the \ion{Si}{4} and \ion{C}{2} profiles.
However,
the redshift speeds
of these events
are generally smaller than
those of the limb-side unipolar events.
The maximum Doppler velocity
of this BP
(Figure \ref{fig:uni_dc})
is $15.6\ {\rm km\ s}^{-1}$.

In Figure \ref{fig:uni_dc},
the single-peaked \ion{Mg}{2} k profiles
(red and purple)
are found near the pore,
which is consistent with
the limb-side unipolar events
(Figure \ref{fig:uni_lm}),
while the blue \ion{Mg}{2} k profile
indicates that
this location is probably covered
by the AFS.
Although not clear,
Figure \ref{fig:uni_dc} shows
an indication
of a slight \ion{Mg}{2} triplet emission.

\subsubsection{Averaged Profiles}

Figure \ref{fig:average} compares
the
profiles
of the \ion{Si}{4}, \ion{Mg}{2},
\ion{C}{2}, and Stokes-V/I
averaged over the areas covered
by the seven mixed-polarity BPs,
five limb-side unipolar BPs,
and five DC-side unipolar BPs.

The mixed-polarity events
(orange)
are characterized
by their enhanced and broadened profiles,
especially of the \ion{Si}{4} and \ion{C}{2} lines.
The peak intensities
of the averaged \ion{Si}{4} and \ion{C}{2} lines
are about 9 times
of the quiet-Sun levels,
while their wings
exceed $\pm 150\ {\rm km\ s}^{-1}$.
The \ion{Mg}{2} k has a double-peaked profile,
which indicates that
the mixed-polarity BPs
are generally located
below the fibrilar canopy
of the AFS.
The \ion{Mg}{2} triplet emission
is no more seen
in this averaged profile
because the triplet emission
is rare and, if occurs,
exists only in the cores
of the BPs.

The clearest feature
of the limb-side unipolar events
(green solid),
which shows a negative polarity
in the Stokes-V/I profile,
is the redshifts
of the \ion{Si}{4} and \ion{C}{2} spectra.
These are caused by
the systematic downflows
at the footpoints
of the AFS.
The peak intensities
of the \ion{Si}{4} and \ion{C}{2} lines
are darker,
only twice and 4 times
of their quiet-Sun values,
respectively.
Another obvious characteristic
is the single-peaked shape
of the \ion{Mg}{2} k spectrum.
Such a profile
is created
probably because
these AFS footpoints are located
above or very close to
the developing pores.
The \ion{Mg}{2} triplet
is seen in emission,
indicating the heating of the lower chromosphere.

Compared to the limb-side events,
the redshifts of the \ion{Si}{4} and \ion{C}{2} lines
of the DC-side, positive unipolar events
(green dashed)
are not very prominent,
while the peak intensities
of these lines
show intermediate values
between the mixed-polarity and
the limb-side unipolar:
six and eight times
the quiet-Sun values.
The \ion{Mg}{2} k has
an asymmetric profile
with the central dip,
probably caused
by the mixture
of the single-peaked profiles
near the pores
and the double-peaked ones
under the AFS.
Similar to the limb-side events,
the \ion{Mg}{2} triplet emission
is seen.

\subsection{Temporal Evolutions}

Figure \ref{fig:lc} plots
the light curves
of several chromospheric
and transition-region lines
for the three representative Ca BPs
introduced in Section \ref{sec:spectral}
(mixed-polarity, limb-side unipolar,
and DC-side unipolar events).
The light curves are
obtained as time-varying intensities
integrated over a $1\arcsec\times 1\arcsec$ box,
centered at the middle of the BPs.
In this figure,
each light curve
is normalized by
its quiet-Sun value.

For the mixed-polarity event
(Figure \ref{fig:lc} top),
the light curves
show drastic enhancement
around 10:45 UT
with a duration of 10 to 15 minutes.
They show
similar evolution profiles
with each other,
a fast rise followed
by an extended decay,
which is highly reminiscent
of the light curves
of H$\alpha$ and soft X-rays
recorded in the solar flares
\citep[e.g.,][]{kan74}.
The most significant enhancement
is of the wavelength-integrated intensity
of the \ion{Si}{4} line,
which increases up to almost 100 times
of its quiet-Sun level.
Similar drastic \ion{Si}{4} enhancements
of more than 10 times
are seen
for most of the mixed-polarity events.

Another feature to note
for the mixed-polarity events
is that the intensity enhancements
are often repetitive
at the same locations.
For the mixed-polarity BP
in Figure \ref{fig:lc},
the preceding event occurs
about 20 minutes before
the main one.

On the other hand,
the light curves
for both limb-side and DC-side unipolar events
are generally less marked.
The evolutions of typical events
in Figure \ref{fig:lc}
(middle and bottom)
are rather steady over time,
and the normalized intensities
become only a few times,
at most 10 times,
of the quiet-Sun levels.
The most enhanced intensities
are again of the \ion{Si}{4} spectra.

\subsection{Statistical Trends}

The statistical trends
for the analyzed 29 Ca BPs
are shown
in Figure \ref{fig:statistics}.
Here,
the Ca intensities
of both types of the BPs
are clearly distinct.
The mixed-polarity events
are generally brighter
than the unipolar ones
with the critical value
being $\sim 1600$ DN.

The maximum area of the Ca BPs
is positively correlated with the intensity.
Most of the mixed-polarity events
have areas ranging
from $0.8\times 10^{6}$
to $4.5\times 10^{6}\ {\rm km}^{2}$,
which is equivalent to
the radii
of 500 to 1200 km
if one assumes circular shapes.
The unipolar BPs are
$\lesssim 2\times 10^{6}\ {\rm km}^{2}$
in area,
or of the radius
of $\lesssim 800\ {\rm km}$.
The Ca lifetime shows,
although more than half
of the measurements
are limited
by the cadence and the durations
of the Ca observation by SOT,
which are
1 minute and 25 minutes,
respectively,
a positive relation
with the Ca intensity.

The plot for the maximum and minimum Doppler velocities,
which is
obtained by simply fitting
a single Gaussian function
to the \ion{Si}{4} profile,
clearly shows the difference
between the two groups.
While the mixed-polarity events
have a wider velocity range
from $-32.4\ {\rm km\ s}^{-1}$ (blueshift)
to $20.4\ {\rm km\ s}^{-1}$ (redshift),
the unipolar ones
are remarkably one-sided
with a range of
$-8.4$ to $35.7\ {\rm km\ s}^{-1}$,
i.e., the downflows (redshifts)
are dominated.
Note that we here use
the single-Gaussian fitting,
which provides
the Doppler velocity
near the core of the spectrum
and may be affected
when the profile contains
of more than two components.

The maximum intensities
of the \ion{Si}{4}, \ion{C}{2},
and \ion{Mg}{2} triplet lines
also show positive relations
with the \ion{Ca}{2} H intensity.
However,
the dynamic range for the \ion{Mg}{2} triplet
is less than one order of magnitudes,
120 to 690 DN,
as opposed to the wider ranges
of the \ion{Si}{4} and \ion{C}{2},
7.3 to 2000 DN
and 19 to 1500 DN,
respectively.

\section{Summary}\label{sec:summary}

In this paper,
we have investigated
various local heating events
in the EFRs
using the co-observation data
obtained by {\it Hinode} and {\it IRIS},
complemented by {\it SDO}.
The main observational results
of our study
are summarized as follows.

The mixed-polarity events
are seen in the central part
of the EFR,
which is characterized
by enhanced and broadened chromospheric
and transition-region spectra.
In particular,
the \ion{Si}{4} line brightens up
significantly to 100 times
of the quiet-Sun intensity.
Sometimes
the \ion{Si}{4} and \ion{C}{2} spectra
spread over
the Doppler speeds
of $\pm 150\ {\rm km\ s}^{-1}$.
The Doppler velocities
at the \ion{Si}{4} line center
have a positional tendency
that the redshifts of up to $20\ {\rm km\ s}^{-1}$
are seen in the DC side
and the blueshifts of up to $30\ {\rm km\ s}^{-1}$
are seen in the limb side
within the BP.
The \ion{Mg}{2} k spectra
often show the
central reversals,
which is caused by
the overlying AFS obscuration,
while sometimes the \ion{Mg}{2} triplet
is seen in emission,
indicating the lower chromospheric heating.
These events
are mostly found
above the PILs
with the bald-patch
(i.e., U-shaped)
magnetic configurations
in the photosphere.
They are larger in size,
longer in lifetime.
The light curves
show flare-like evolutions
with durations of 10 -- 15 minutes
and the brightenings recur
with periods
of about 20 minutes.

The unipolar events
are mostly found
in the peripheral regions
of the EFR,
i.e., the footpoints
of the AFS.
These events are weaker
in intensities
with the maximum \ion{Si}{4} intensity
being at most 10 times
of the quiet-Sun level.
The \ion{Si}{4} spectra show
systematic downflows
with the maximum Doppler speed
of $\sim 40\ {\rm km\ s}^{-1}$
at the line center.
The redshift velocity
is generally faster
for the limb-side events
than the DC-side events.
If the BP is close to
or above a pore,
the \ion{Mg}{2} k spectrum
reveals a single-peaked profile
with no absorption core,
similar to the spectra
found in the plage regions.
The \ion{Mg}{2} triplet is seen
in emission occasionally.
They are smaller in size,
shorter in lifetime,
and the evolution is steady.

\section{Discussion}\label{sec:discussion}

The observational results
summarized in Section \ref{sec:summary}
lend support to the physical picture
illustrated in Figure \ref{fig:illust}.
In EFRs,
the magnetic fields appear
at the photosphere
as $\Omega$-shaped or undular field lines
and evolve through episodes
of merging and cancellations
of fragmentary magnetic elements.
The AFS is seen above,
which may represent
the ascending magnetic fields
in the higher altitudes.

\begin{figure}
  \begin{center}
    \includegraphics[width=165mm]{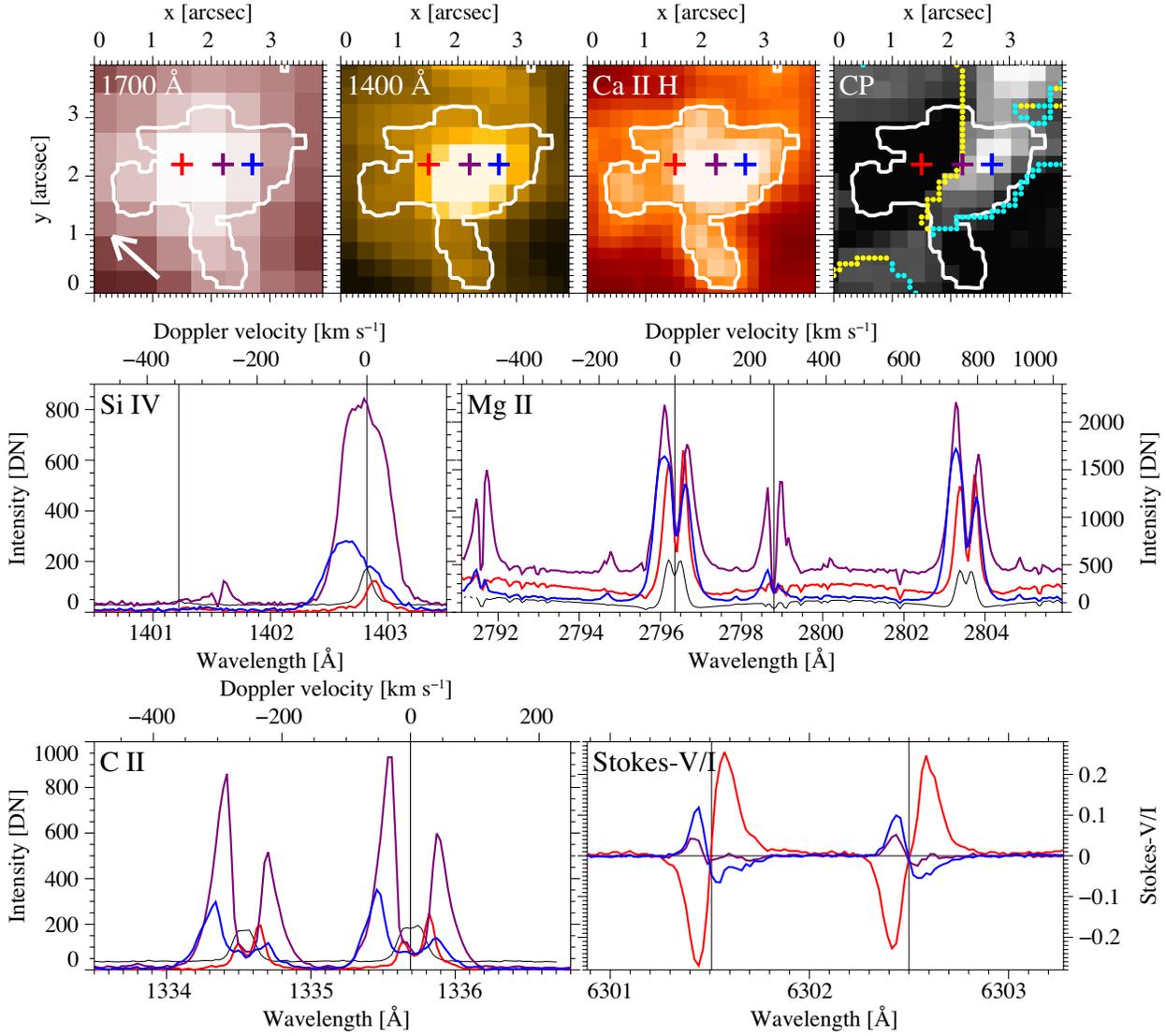}
  \end{center}
  \caption{
    Spectral and polarimetric profiles
    of the sample mixed-polarity Ca BP
    (event A)
    located at
    $(X, Y)=(336\arcsec, -274\arcsec)$
    in Figure \ref{fig:eb_sp}(b).
    Four panels at the top
    are the filtergrams
    of AIA 1700 {\AA},
    {\it IRIS} SJI 1400 {\AA},
    SOT \ion{Ca}{2} H,
    and the SOT SP circular polarization map.
    The white contour delineates
    the Ca BPs,
    while the white arrow
    in the 1700 {\AA} image
    shows the direction of the DC.
    The dots in the circular polarization map
    indicate the locations
    of the PIL,
    where yellow and turquoise
    show that the magnetic fields
    have ``dip'' (i.e, bald-patch)
    and ``top'' structures,
    respectively
    (see text for details).
    Four bottom panels provide
    the {\it IRIS} spectra
    of \ion{Si}{4}, \ion{C}{2}, and \ion{Mg}{2},
    and the SOT SP Stokes-V/I.
    The profiles of different colors
    (red, purple, blue) are sampled at
    three location
    indicated by the $+$ signs
    of corresponding colors
    in the top four images,
    while the black profiles
    are the averaged quiet-Sun levels
    obtained from the same data sets.
    In the \ion{Si}{4} and \ion{C}{2} panels,
    the quiet-Sun profiles
    are multiplied by factors of 10.
    The vertical lines show
    the main lines
    and some blends,
    which are
    \ion{O}{4} 1401.2 {\AA},
    \ion{Si}{4} 1402.8 {\AA},
    \ion{Mg}{2} k 2796.4 {\AA},
    \ion{Mg}{2} triplet 2798.8 {\AA},
    \ion{C}{2} 1335.7 {\AA},
    \ion{Fe}{1} 6301.5 {\AA},
    and \ion{Fe}{1} 6302.5 {\AA}.
  }
  \label{fig:mix}
\end{figure}

\begin{figure}
  \begin{center}
    \includegraphics[width=165mm]{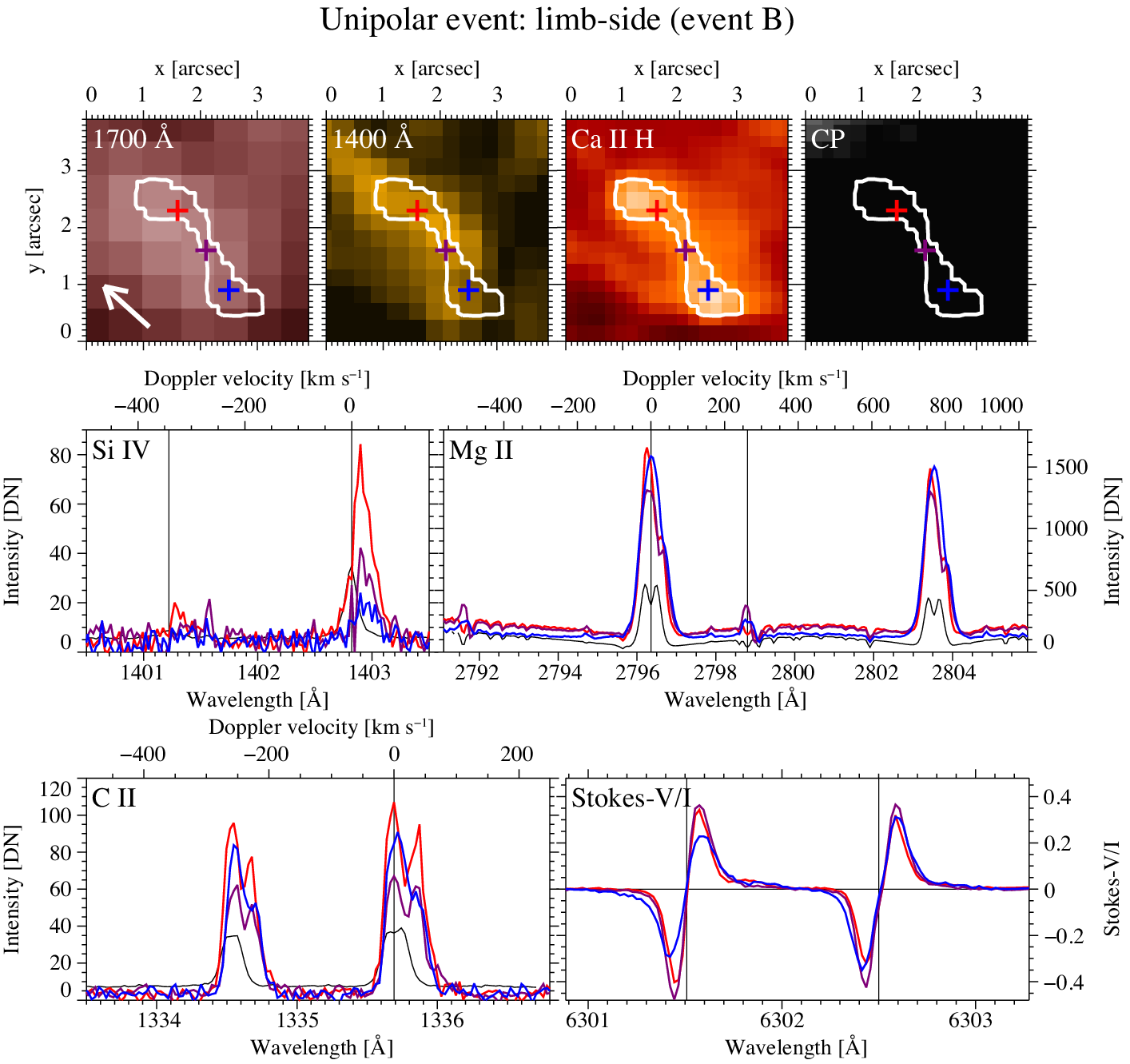}
  \end{center}
  \caption{
    The same as Figure \ref{fig:mix}
    but for the limb-side unipolar event
    (event B)
    at $(X, Y)=(350\arcsec, -270\arcsec)$
    in Figure \ref{fig:eb_sp}(b).
    The intensity scales of the top panels
    are identical to those of Figure \ref{fig:mix}.
    In the \ion{Si}{4} and \ion{C}{2} panels,
    the quiet-Sun profiles
    are multiplied by factors of 2.
  }
  \label{fig:uni_lm}
\end{figure}

\begin{figure}
  \begin{center}
    \includegraphics[width=165mm]{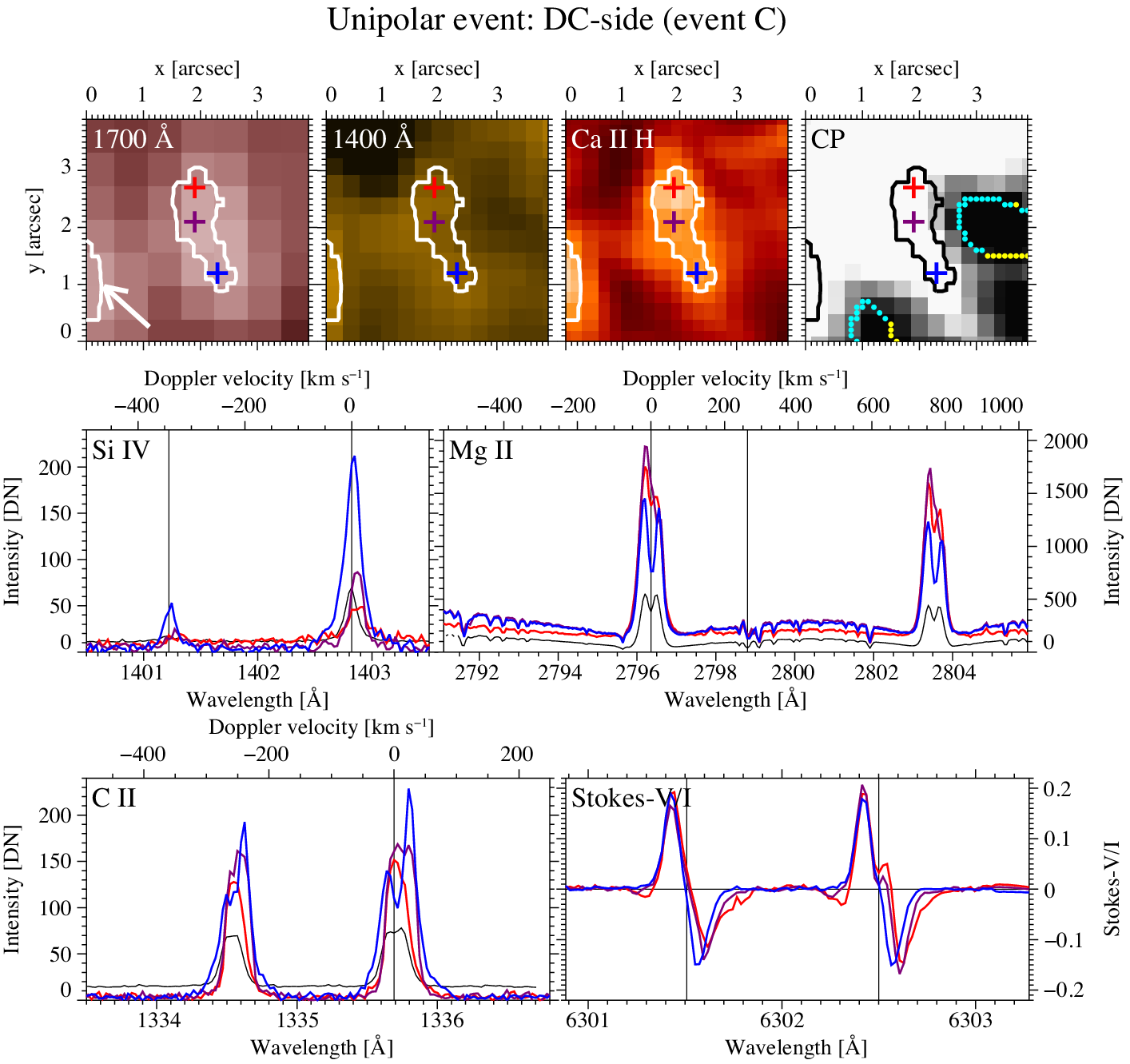}
  \end{center}
  \caption{
    The same as Figure \ref{fig:mix}
    but for the DC-side unipolar event
    (event C)
    at $(X, Y)=(328\arcsec, -266\arcsec)$
    in Figure \ref{fig:eb_sp}(b).
    The intensity scales of the top panels
    are identical to those of Figure \ref{fig:mix}.
    In the \ion{Si}{4} and \ion{C}{2} panels,
    the quiet-Sun profiles
    are multiplied by factors of 4.
  }
  \label{fig:uni_dc}
\end{figure}

\begin{figure}
  \begin{center}
    \includegraphics[width=170mm]{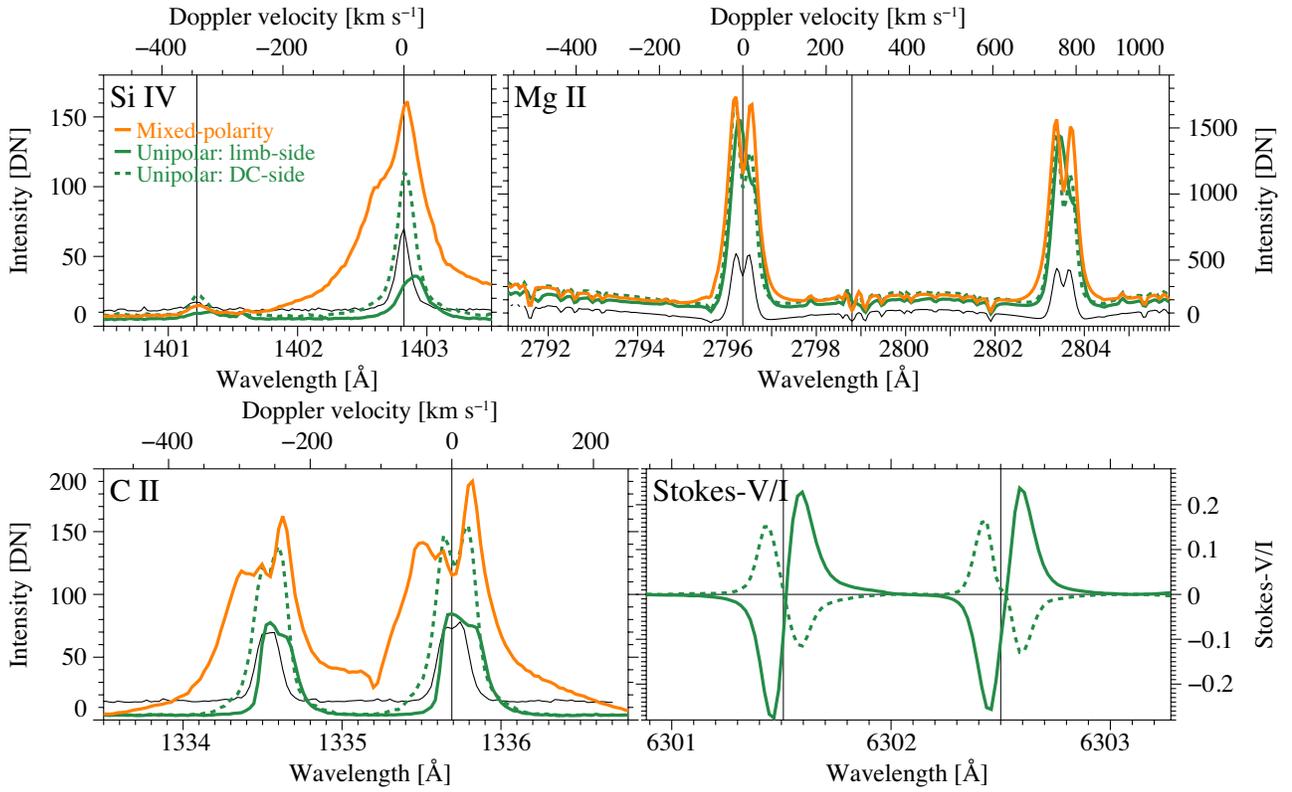}
  \end{center}
  \caption{
    Averaged profiles of 
    the \ion{Si}{4}, \ion{Mg}{2}, \ion{C}{2},
    and Stokes-V/I
    for the seven mixed-polarity events
    (orange),
    five limb-side unipolar events
    (green solid),
    and five DC-side unipolar events
    (green dashed).
    The black curves and vertical lines
    are identical to those
    in Figure \ref{fig:mix}.
    In the \ion{Si}{4} and \ion{C}{2} panels,
    the quiet-Sun profiles
    are multiplied by factors of 4.
  }
  \label{fig:average}
\end{figure}

\begin{figure}
  \begin{center}
    \includegraphics[width=70mm]{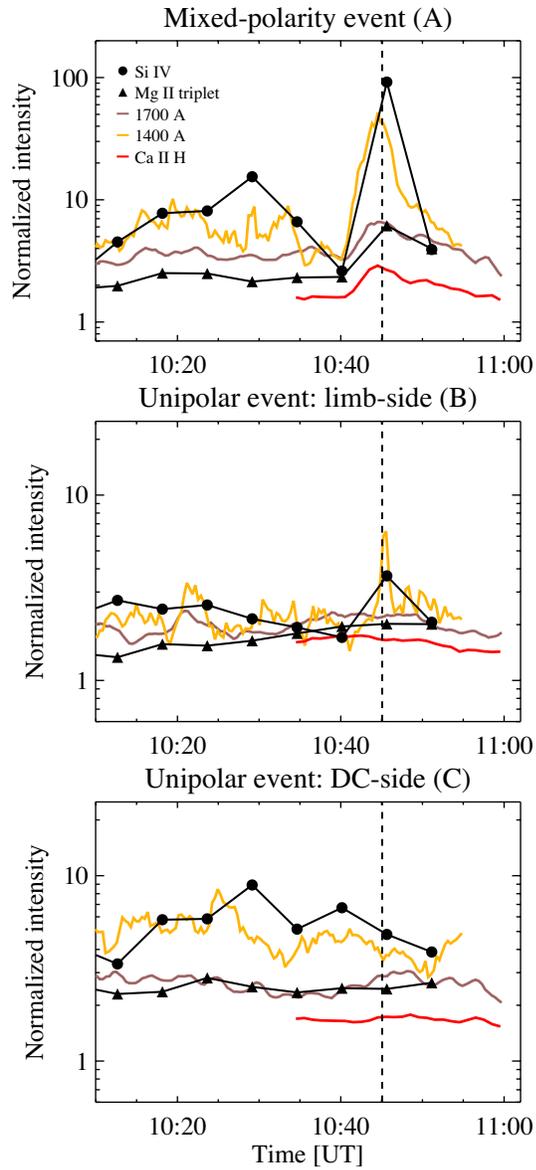}
  \end{center}
  \caption{
    Temporal evolutions
    of the chromospheric and transition-region line intensities
    integrated over the $1\arcsec\times 1\arcsec$ box
    at the center of the BPs
    for the three representative events
    in Section \ref{sec:spectral}.
    The plotted intensities
    of the \ion{Si}{4} and \ion{Mg}{2} triplet lines
    are the intensities
    integrated over the spectral ranges of
    $\pm 0.5\ {\rm \AA}$ and $\pm 0.75\ {\rm \AA}$
    from the line centers,
    respectively.
    All the intensities
    are normalized
    by the quiet-Sun values.
    Vertical dashed lines
    indicate the timing
    of the SOT SP scan.
  }
  \label{fig:lc}
\end{figure}

\begin{figure}
  \begin{center}
    \includegraphics[width=150mm]{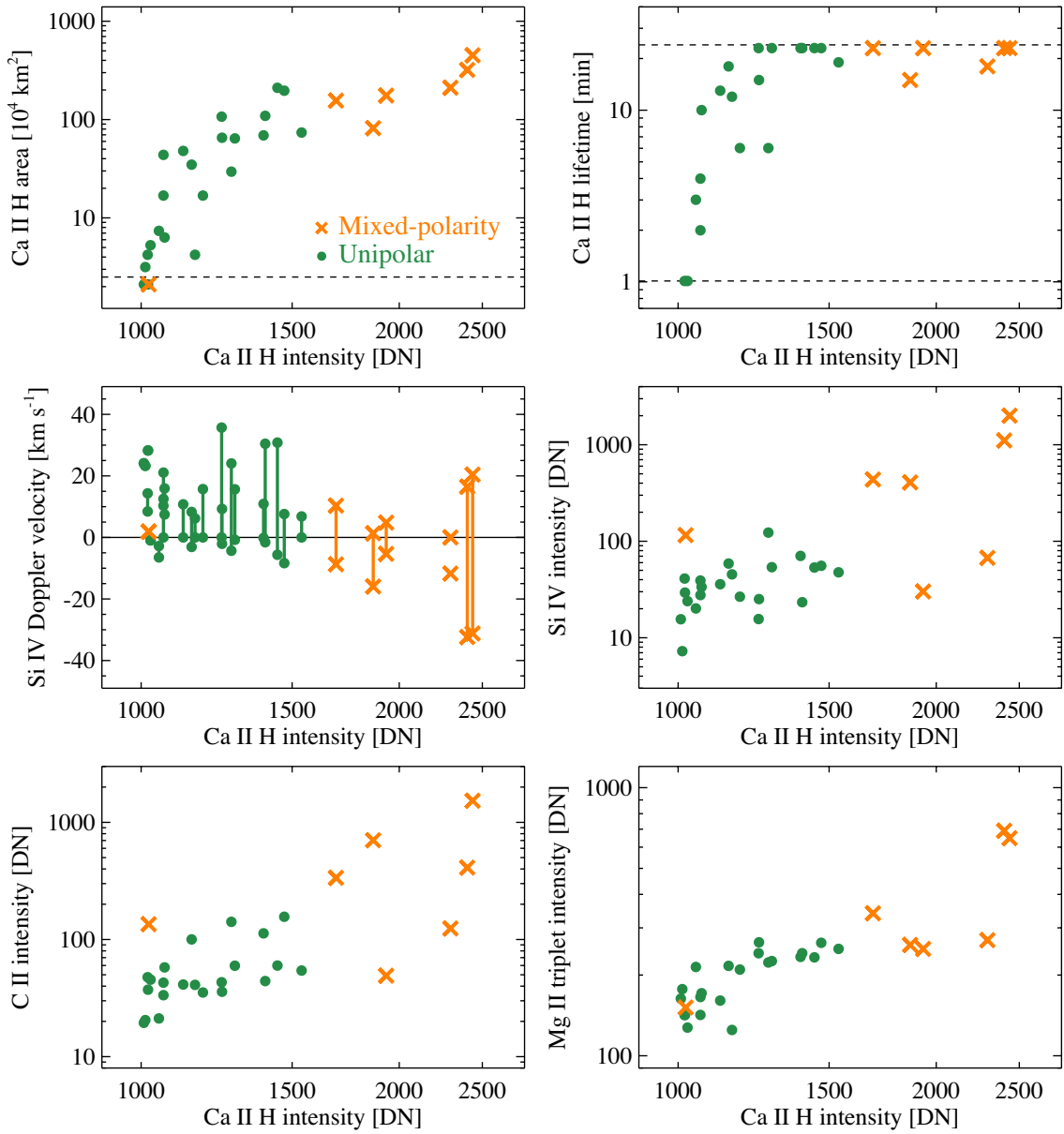}
  \end{center}
  \caption{
    From top left to bottom right,
    scatter plots showing
    the maximum area
    and lifetime (or duration)
    measured in \ion{Ca}{2} H,
    the maximum and minimum Doppler velocities
    measured from the \ion{Si}{4} spectra,
    the maximum intensities
    for the \ion{Si}{4}, \ion{C}{2}, and \ion{Mg}{2} triplet lines
    for all the mixed-polarity BPs
    (seven events: orange $\times$-marks)
    and unipolar BPs
    (22 events: green dots)
    as a function of \ion{Ca}{2} H intensity.
  }
  \label{fig:statistics}
\end{figure}

\begin{figure}
  \begin{center}
    \includegraphics[width=160mm]{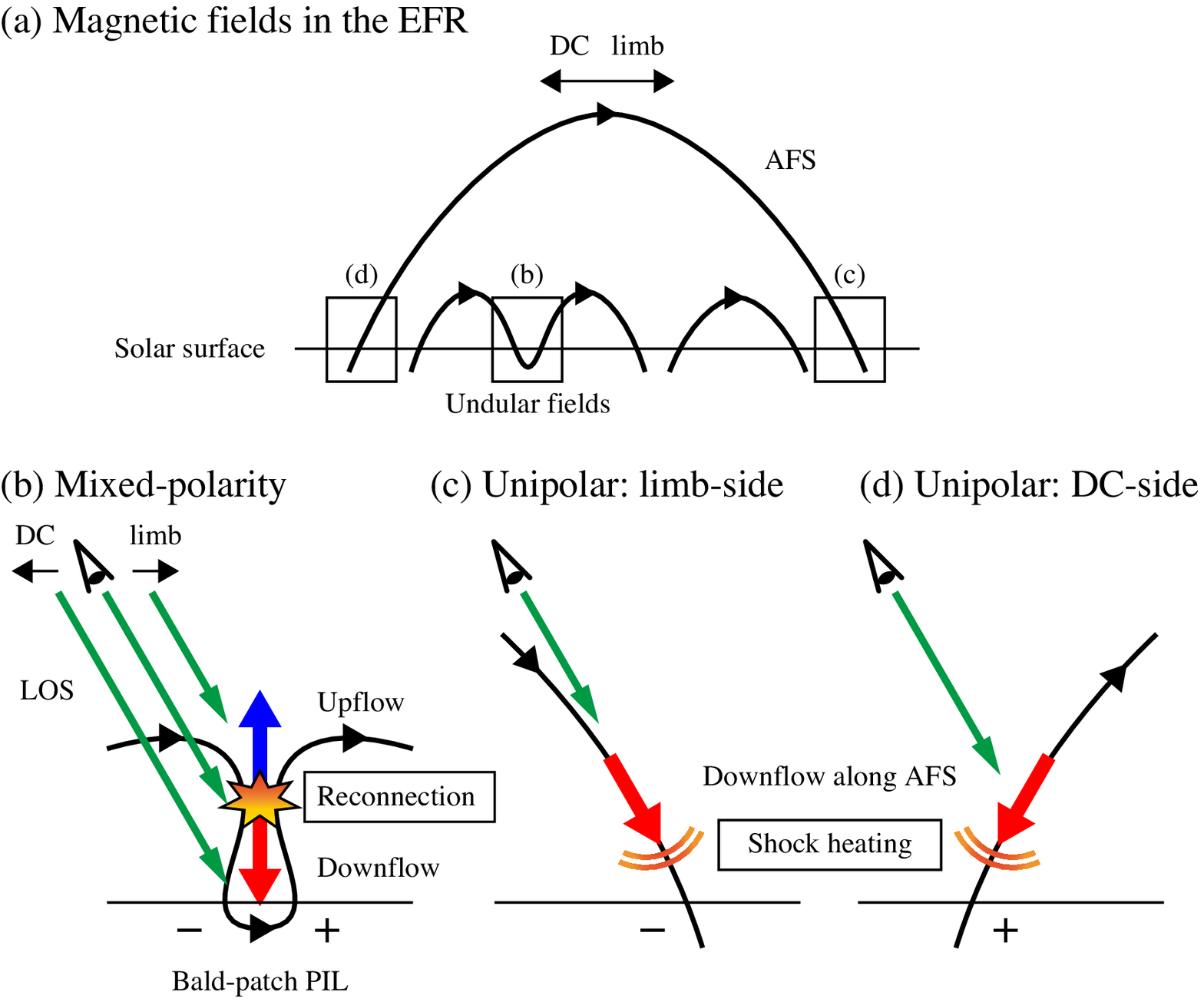}
  \end{center}
  \caption{
    Schematic illustration
    summarizing the various local heating events
    in the EFRs.
    (a) Overview of the magnetic fields
    in the EFRs.
    Thick curve lines
    and a thin horizontal line
    indicate magnetic field lines
    and the solar surface.
    Three boxes show
    the regions
    corresponding to panels (b) to (d).
    (b) Mixed-polarity events,
    where the BP may represent
    the magnetic reconnection
    between the cancelling opposite magnetic polarities.
    The field line configuration
    is similar to
    Figure 12 of \citet{geo02}.
    Red and blue arrows indicate
    bi-directional jets,
    while the green arrows present
    the different LOSs.
    The $+$ and $-$ signs show
    the positive and negative polarities,
    respectively.
    (c and d) Limb-side and DC-side
    unipolar events,
    where the BP may be
    due to the shock heating
    caused by the fast downflows
    along the AFS.
    Orange curves represent
    the local heating.
  }
  \label{fig:illust}
\end{figure}

The observed mixed-polarity Ca BPs
may indicate
magnetic reconnection
between the cancelling positive and negative polarities,
as shown in Figure \ref{fig:illust}(b).
Although we have no H$\alpha$ observation
for our events,
these BPs should be
closely related to
the classical Ellerman bombs
\citep{ell17}.
The striking consistencies
may be found
if one compares
Figure \ref{fig:mix} of this paper
and Figure 4 of \citet{vis15b},
in which they show
the {\it IRIS} and H$\alpha$ spectra
of a textbook Ellerman bomb.
Recently,
\citet{tia16} and \citet{gru16}
demonstrated that
the \ion{Mg}{2} h and k lines
of Ellerman bombs
show intense brightening in the wings
and no significant enhancement in the core,
similar to the H$\alpha$ profiles.
The \ion{Mg}{2} spectra
in Figure \ref{fig:mix}
agree with this trend,
which further supports
that our events are likely
Ellerman bombs.

Recent high-resolution observations
indicate that
the Ellerman bombs are
the photospheric reconnection
with flame-like structures
\citep{wat11,vis13,rut13}.
Such flames may not be resolved
in our observations
with a minimum pixel size of
$0\farcs22$ (\ion{Ca}{2} H).
Also, especially in 1700 {\AA},
they may be blurred
by scattering
\citep{rut16}.

The red and blueshifts
observed in the mixed-polarity BPs
can be interpreted
as the bi-directional jets
from the reconnection sites.
The DC- and limb-ward dependence
of the Doppler velocity
may be due to the viewing angle.
As the target EFR
of this study
is located
away from the DC
by $26^{\circ}$ -- $27^{\circ}$,
the slanted viewing angle
may cause the sampling of upflows
in the limb-side
(or upper parts)
and downflows
in the DC-side
(lower parts).
Perhaps one might expect
the siphon flow
along the small $\Omega$-loop
for explaining the red and blueshifts.
However,
the dominance of
the bald-patch PILs
for the mixed-polarity BPs
favors the U-loop cancellation theory
\citep{geo02,par04,wat08}.

The unipolar BPs
at the periphery
of the EFR
may be caused by the downflows
(Figures \ref{fig:illust}(c) and (d)).
\citet{bru69} found
the fast downdrafts
of $\sim 50\ {\rm km\ s}^{-1}$
(in H$\alpha$)
along the field lines of the AFS
at both footpoints.
Using magnetohydrodynamic (MHD) simulations,
\citet{shi89} modeled the EFRs
and suggested that
the bright plages
near the AFS footpoints
are caused by the shocks,
which is produced by
the fast downflows
of 30 to $50\ {\rm km\ s}^{-1}$
along the AFS
entering the high-density lower atmosphere.
The above observations and simulations
point to the possibility that
the unipolar BPs
in the present study
are also due to the downflows.

If we simply assume
a CHIANTI-based
(i.e., coronal-equilibrium)
\ion{Si}{4} formation temperature
of $\log{T\, [{\rm K}]}=4.8$ -- $4.9$
\citep{dep14,pet14},
the local adiabatic sound speed is
$\gtrsim 40\ {\rm km\ s}^{-1}$.
However,
it is more likely that
the plasma at the AFS footpoints
is denser
and that the \ion{Si}{4} temperature
is cooler
than these values.
% high density + high temp => Saha-Boltzmann
Since \citet{rut16} suggests
$15,000$ to $20,000\ {\rm K}$
for the \ion{Si}{4} temperature
for the Ellerman bombs
(the heating of low-chromospheric to mid-photospheric plasma),
let us assume,
say, $40,000\ {\rm K}$
as the temperature.
Then, we obtain
the sound speed
of $\sim 30\ {\rm km\ s}^{-1}$.
Therefore,
there is a good chance that
the downflows
observed above the unipolar regions,
whose Doppler velocity
is up to $35.7\ {\rm km\ s}^{-1}$,
are comparable to
or exceed
the local sound speed.
Although we did not directly detect
the shock fronts,
considering
that the \ion{Mg}{2} lines
are not strongly Doppler-shifted
(Section \ref{sec:lue})
and that the typical sound speed
in the photosphere
is by far smaller,
there may be shocks,
or at least strong compressions
and condensations of the material,
which locally heat the atmosphere.

The differences of the observational characteristics
between the limb-side and DC-side events,
especially of the Doppler velocities,
may be due to the viewing angle:
compare Figures \ref{fig:illust}(c) and (d).

Perhaps this scenario
has a relation with
the transient supersonic downflows
and associated heating events
found above the sunspots
by \citet{kle14} and \citet{tia14}.
However, their downflows are
by far faster,
exceeding $200\ {\rm km\ s}^{-1}$,
and probably are the material
falling higher from the coronal heights
such as coronal rain,
as opposed to our steady, much slower downflows
originated from the AFS.

The heating mechanisms
suggested in this section
explain the observations
without inconsistencies
and are, although not perfectly,
supported by
MHD simulations
\citep[e.g.,][]{iso07,tor09,che10,shi89}.
However,
especially for the heating
in the unipolar regions,
we cannot rule out
other possibilities
such as magnetic reconnection
that we cannot resolve,
or the component reconnection
between the field lines
of colliding magnetic patches
of the same polarity
(see, e.g., Figure 12c of \citealt{geo02}).
Therefore,
it is of necessity
to numerically model these mechanisms
with taking into account
the radiative transfer
and examine the proposed scenarios.

%\clearpage

\acknowledgments

The authors are grateful
to the anonymous referee
for improving the manuscript.
The authors thank ISSI Bern
for the support to the team
``Solar UV bursts --- a new insight to magnetic reconnection.''
S.T. would like to thank Dr. Robert Rutten,
who visited NAOJ
and gave a lecture
on the solar spectral formation,
for fruitful discussion.
Data are courtesy of the science teams
of {\it Hinode}, {\it IRIS}, and {\it SDO}.
{\it Hinode} is a Japanese mission
developed and launched by ISAS/JAXA,
with NAOJ as domestic partner
and NASA and STFC (UK) as international partners.
It is operated by these agencies
in cooperation with ESA and NSC (Norway).
{\it IRIS} is a NASA small explorer mission
developed and operated by LMSAL
with mission operations executed at NASA Ames Research center
and major contributions to downlink communications funded by ESA
and the Norwegian Space Centre.
HMI and AIA are instruments on board {\it SDO},
a mission for NASA's Living With a Star program.
This work was carried out on the Solar Data Analysis System
operated by the Astronomy Data Center
in cooperation with the Hinode Science Center of NAOJ.
This work was supported by
JSPS KAKENHI Grant Numbers JP16K17671 (PI: S. Toriumi),
JP15H05814 (PI: K. Ichimoto),
and JP25220703 (PI: S. Tsuneta).
MCMC acknowledges support by
NASA contracts NNG09FA40C ({\it IRIS}),
NNG04EA00C ({\it SDO}/AIA)
and NNM07AA01C ({\it Hinode}/SOT),
and grant NNX14AI14G (Heliophysics Grand Challenges Research).


%% for the paper.  Each keyword is check against the master list during
%% copy editing.  Individual instruments can be provided in parentheses,
%% after the keyword, but they are not verified.

\vspace{5mm}

\clearpage

\clearpage

\clearpage

\clearpage

\clearpage

\clearpage

\clearpage

\clearpage

\clearpage

\clearpage
\appendix

\section{PILs of Mixed-polarity Events}\label{app:bald}

\begin{figure}
  \begin{center}
    \includegraphics[width=120mm]{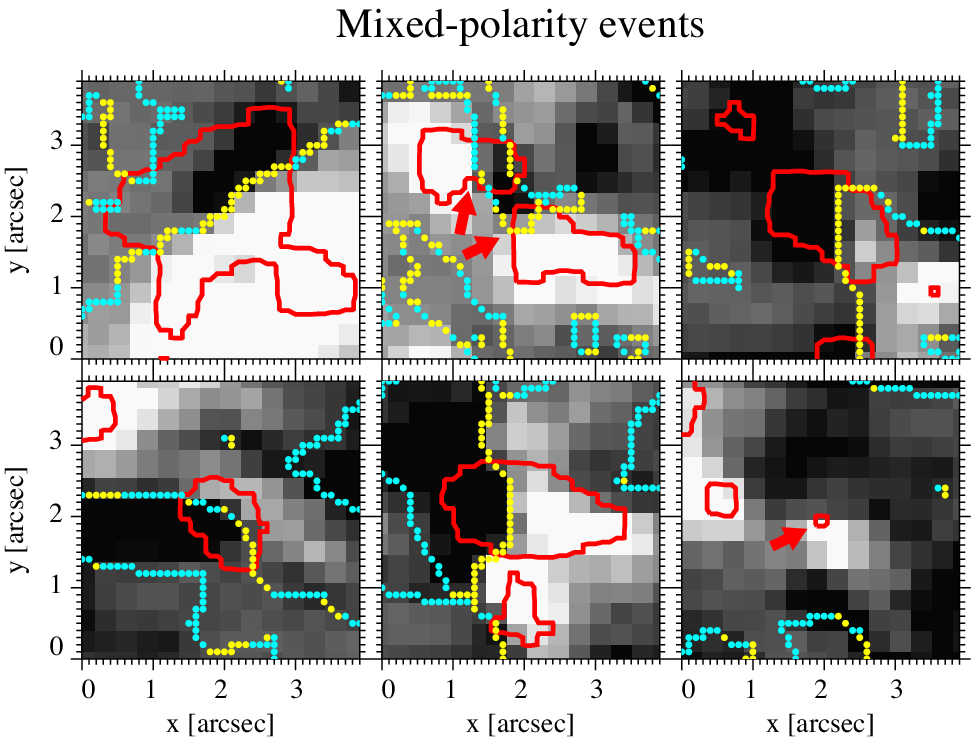}
  \end{center}
  \caption{
    Six remaining mixed-polarity events
    located at (from top left to bottom right)
    $(X, Y)=(335\arcsec, -280\arcsec)$,
    $(328\arcsec, -274\arcsec)$,
    $(336\arcsec, -272\arcsec)$,
    $(336\arcsec, -265\arcsec)$,
    $(340\arcsec, -265\arcsec)$,
    and $(333\arcsec, -267\arcsec)$
    in Figure \ref{fig:eb_sp}(b).
    Each panel shows
    the SP circular polarization map
    (black-white),
    \ion{Ca}{2} H contour (red)
    defining the Ca BPs,
    and locations of PILs (dots),
    where yellow and turquoise
    indicate that
    the magnetic fields have ``dip''
    (i.e, bald-patch)
    and ``top'' structures,
    respectively.
    In the top-middle panel,
    a single event is separated
    into two patches
    at the moment of SOT SP scan.
  }
  \label{fig:bald}
\end{figure}

Figure \ref{fig:bald} summarizes
the six remaining mixed-polarity events.
Here,
except for the two events
(shown with red arrows),
the PILs are dominated
by the field lines
with bald-patch configurations.
Combined with the event
introduced in Section \ref{sec:mix},
one can see that
five out of seven total mixed-polarity BPs
(71\%)
reveal bald-patch PILs.
The event shown in the top middle panel
of Figure \ref{fig:bald}
is separated in two patches
with both ``dip''-dominated and ``top''-dominated PILs
(red arrows),
which may also support
the importance of bald-patch PILs
for the mixed-polarity events.

\end{document}